\documentclass{article}
\usepackage{amsmath, amssymb, amsfonts}
\title{ The membrane paradigm for a dynamic wormhole} 
\author{Hristu Culetu, \\Ovidius University, Dept.of Physics, \\B-dul Mamaia 124, 900527 Constanta, Romania, \\e-mail : hculetu@yahoo.com}

\begin{document}
\numberwithin{equation}{section}
\pagenumbering{arabic}
\maketitle
\newcommand{\fv}{\boldsymbol{f}}
\newcommand{\tv}{\boldsymbol{t}}
\newcommand{\gv}{\boldsymbol{g}}
\newcommand{\OV}{\boldsymbol{O}}
\newcommand{\wv}{\boldsymbol{w}}
\newcommand{\WV}{\boldsymbol{W}}
\newcommand{\NV}{\boldsymbol{N}}
\newcommand{\hv}{\boldsymbol{h}}
\newcommand{\yv}{\boldsymbol{y}}
\newcommand{\RE}{\textrm{Re}}
\newcommand{\IM}{\textrm{Im}}
\newcommand{\rot}{\textrm{rot}}
\newcommand{\dv}{\boldsymbol{d}}
\newcommand{\grad}{\textrm{grad}}
\newcommand{\Tr}{\textrm{Tr}}
\newcommand{\ua}{\uparrow}
\newcommand{\da}{\downarrow}
\newcommand{\ct}{\textrm{const}}
\newcommand{\xv}{\boldsymbol{x}}
\newcommand{\mv}{\boldsymbol{m}}
\newcommand{\rv}{\boldsymbol{r}}
\newcommand{\kv}{\boldsymbol{k}}
\newcommand{\VE}{\boldsymbol{V}}
\newcommand{\sv}{\boldsymbol{s}}
\newcommand{\RV}{\boldsymbol{R}}
\newcommand{\pv}{\boldsymbol{p}}
\newcommand{\PV}{\boldsymbol{P}}
\newcommand{\EV}{\boldsymbol{E}}
\newcommand{\DV}{\boldsymbol{D}}
\newcommand{\BV}{\boldsymbol{B}}
\newcommand{\HV}{\boldsymbol{H}}
\newcommand{\MV}{\boldsymbol{M}}
\newcommand{\be}{\begin{equation}}
\newcommand{\ee}{\end{equation}}
\newcommand{\ba}{\begin{eqnarray}}
\newcommand{\ea}{\end{eqnarray}}
\newcommand{\bq}{\begin{eqnarray*}}
\newcommand{\eq}{\end{eqnarray*}}
\newcommand{\pa}{\partial}
\newcommand{\f}{\frac}
\newcommand{\FV}{\boldsymbol{F}}
\newcommand{\ve}{\boldsymbol{v}}
\newcommand{\AV}{\boldsymbol{A}}
\newcommand{\jv}{\boldsymbol{j}}
\newcommand{\LV}{\boldsymbol{L}}
\newcommand{\SV}{\boldsymbol{S}}
\newcommand{\av}{\boldsymbol{a}}
\newcommand{\qv}{\boldsymbol{q}}
\newcommand{\QV}{\boldsymbol{Q}}
\newcommand{\ev}{\boldsymbol{e}}
\newcommand{\uv}{\boldsymbol{u}}
\newcommand{\KV}{\boldsymbol{K}}
\newcommand{\ro}{\boldsymbol{\rho}}
\newcommand{\si}{\boldsymbol{\sigma}}
\newcommand{\thv}{\boldsymbol{\theta}}
\newcommand{\bv}{\boldsymbol{b}}
\newcommand{\JV}{\boldsymbol{J}}
\newcommand{\nv}{\boldsymbol{n}}
\newcommand{\lv}{\boldsymbol{l}}
\newcommand{\om}{\boldsymbol{\omega}}
\newcommand{\Om}{\boldsymbol{\Omega}}
\newcommand{\Piv}{\boldsymbol{\Pi}}
\newcommand{\UV}{\boldsymbol{U}}
\newcommand{\iv}{\boldsymbol{i}}
\newcommand{\nuv}{\boldsymbol{\nu}}
\newcommand{\muv}{\boldsymbol{\mu}}
\newcommand{\lm}{\boldsymbol{\lambda}}
\newcommand{\Lm}{\boldsymbol{\Lambda}}
\newcommand{\opsi}{\overline{\psi}}
\renewcommand{\tan}{\textrm{tg}}
\renewcommand{\cot}{\textrm{ctg}}
\renewcommand{\sinh}{\textrm{sh}}
\renewcommand{\cosh}{\textrm{ch}}
\renewcommand{\tanh}{\textrm{th}}
\renewcommand{\coth}{\textrm{cth}}

\begin{abstract}
 A classical membrane model is developed for a dynamic wormhole. Its throat expands hyperbolically with an acceleration proportional to the surface energy density $\sigma$. The model leads to a negative bulk viscosity coefficient $\zeta = - 1/6 \pi$ on the throat, expressing the instability against perturbations (expansions or contractions).
\end{abstract}


\section{Introduction}
 Much attention has been paid in the last decades to wormhole physics, especially to traversable Lorentzian wormholes \cite{MT} \cite{MV} \cite{PV}. They may in principle be used by humans to travel between universes or to distant parts of the same universe. Violations of the weak energy condition (WEC) and null energy condition (NEC)  \cite{HV} \cite{HC1} are supposed to occur at the throat of a traversable wormhole.
 
 Visser \cite{MV} and Poisson and Visser \cite{PV} conjectured that all ''exotic'' matter is confined to a thin boundary layer between universes (the wormhole throat). Their model is constructed by surgically grafting two Schwarzschild spacetimes together so that no event horizon is permitted to form. It is achieved by the standard ''cut - and - paste'' construction \cite{MV} : take two copies $(\pm)$ of Schwarzschild's spacetime and remove from each manifold the regions $r_{\pm} \leq R,~ R>2m$, where $R$ is a constant and $m$ - the black hole mass. The resulting manifold have the boundary $\partial\Omega_{\pm} = \left\{r_{\pm} = R,~ R>2m\right\}$. The manifold obtained by identifying the two timelike hypersurfaces $(\partial\Omega_{+} = \partial\Omega_{-} \equiv \partial \Omega)$ possesses two asymptotically flat regions connected by a traversable wormhole. Its throat is the common boundary $\partial\Omega$.\\
 From now on we use geometrical units $G = c = 1$ and the positive signature (-, +, +, +).
 
 \section{The dynamic wormhole}
 According to Birkhoff's theorem, the exterior geometry remains static, in spite of the time dependence of $R(t)$. The velocity 4 - vector of the expanding throat is given by \cite{MV} 
 \begin{equation}
 u^{\alpha} = ( \frac{dt}{d\tau}, \frac{dR}{d\tau}, 0, 0) = ( \frac{\sqrt{1-\frac{2m}{R}+ \dot{R}^{2}}}{1-\frac{2m}{R}}, \dot{R}, 0, 0)
 \label{2.1}
 \end{equation}
 with $\dot{R} = dR/d\tau$, where $\tau$ is the proper time along the throat. We made use of the fact that the Schwarzschild radial coordinate $r = R(t)$ and, therefore, the components of $u^{\alpha}$ depend on time only (spherical symmetry is supposed to hold). 
 
 The normal $n^{\alpha}$ to the surface $\partial\Omega$ observes the relations $n^{\alpha} n_{\alpha} =1$ and $n_{\alpha} u^{\alpha} = 0$. Hence, one obtains 
 \begin{equation}
 n^{\alpha} = ( \frac{\dot{R}}{1-\frac{2m}{R}}, \sqrt{1-\frac{2m}{R}+ \dot{R}^{2}}, 0, 0)
 \label{2.2}
 \end{equation}
 Therefore, the components of the extrinsic curvature of the throat look as \cite{MV} \cite{IS} \cite{CL} 
 \begin{equation}
 \bar{K}_{\theta}^{\theta} = \bar{K}_{\phi}^{\phi} = \frac{2}{R} \sqrt{1-\frac{2m}{R} + \dot{R}^{2}}
 \label{2.3}
 \end{equation}
 and 
 \begin{equation}
 \bar{K}_{\tau}^{\tau} \equiv \bar{a} = \frac{2(\ddot{R}+ \frac{m}{R^{2}})}{\sqrt{1-\frac{2m}{R}+ \dot{R}^{2}}}
 \label{2.4}
 \end{equation}
 where $\bar{K}^{i}_{j} \equiv K^{i}_{j,+} - K^{i}_{j,-} (i,j = \tau, \theta, \phi)$ represents the jump of the extrinsic curvature tensor when the throat is crossed. In addition, $a^{\alpha} = u^{\beta} \nabla_{\beta} u^{\alpha}$ is the 4 - acceleration of the throat and $a^{2} = a_{\alpha} a^{\alpha}$. Using now the expression of the surface stress tensor $S^{i}_{j}$ in terms of the jump in the second fundamental form 
 \begin{equation}
 -8 \pi S^{i}_{j} = \bar{K}^{i}_{j} - \bar{K} \delta^{i}_{j},
 \label{2.5}
 \end{equation}
 obtained from Einstein's equations, one yields
 \begin{equation}
 -2\pi R~ \sigma = \sqrt{1-\frac{2m}{R}+ \dot{R}^{2}}
 \label{2.6}
 \end{equation}
 and
 \begin{equation}
 4\pi R~ p = \frac{1-\frac{m}{R}+ \dot{R}^{2}+ R \ddot{R}}{\sqrt{1-\frac{2m}{R}+ \dot{R}^{2}}}
 \label{2.7}
 \end{equation}
 Let us note that we have $K^{i}_{j,+} = - K^{i}_{j,-} $, due to the spherical symmetry. Therefore, even though Ipser and Sikivie \cite{IS} analyzed the problem of planar domain walls, their case possessses mirror symmetry , just as our case is symmetric w.r.t. $(+) \rightarrow (-)$ change (the throat plays the role of the planar wall). That is not surprising if we remind that, in Ipser and Sikivie notations, the location of the domain wall is given by 
 \begin{equation}
 x^{2} + y^{2} + z^{2} - t^{2} = ( 2 \pi \sigma)^{-1}
 \label{2.8}
 \end{equation}
 in Minkowskian coordinates ($\sigma$ is here positive). The planar domain wall is not a plane at all but rather an accelerating sphere, with constant outward acceleration $2 \pi \sigma$.
 
 We observe that the surface energy density $\sigma$ in (2.6) is negative, as expected (the WEC is violated on the wormhole throat. That is a consequence of the Raychaudhuri equation \cite{PV}). 
 
 \section{The throat equation of motion}
 Let us suppose that the equation of state on the throat is given by the classical membrane
 \begin{equation}
 \sigma + p = 0 .
 \label{3.1}
 \end{equation}
 Eq. (3.1) is, in fact, the definition of a domain wall \cite{IS} , which has repulsive gravitational properties. Combining (2.6) - (2.7) with (3.1) we find that $\sigma = const.$ and the equation for $R(\tau)$ appears as
 \begin{equation}
 R \ddot{R} - \dot{R}^{2} + \frac{3m}{R} -1 = 0.
 \label{3.2}
 \end{equation}
 If $R$ is much greater than the Schwarzschild radius associated to the mass $m$, we may neglect the third term on the l.h.s. of (3.2) and obtain
 \begin{equation}
 R \ddot{R} - \dot{R}^{2} = 1,
 \label{3.3}
 \end{equation}
 with the general solutions
 \begin{equation}
 R(\tau) = \frac{1}{4} e^{-e^{C_{1}} \tau - C_{1} -C_{2} e^{C_{1}}} \left[4 + e^{2e^{C_{1}}(\tau + C_{2})}
 \right] 
 \label{3.4}
 \end{equation}
 and
 \begin{equation}
  R(\tau) = \frac{1}{4} e^{-e^{C_{1}} \tau - C_{1} -C_{2} e^{C_{1}}} \left[1 + 4 e^{2e^{C_{1}}(\tau + C_{2})}
 \right]
 \label{3.5}
 \end{equation}
 where $C_{1}$ and $C_{2}$ are integration constants. Taking $C_{1} = 0$ and rearranging the terms,  it is easy to show that $R(\tau)$ can be written in the form
 \begin{equation}
 R(\tau) = \frac{1}{g} coshg\tau
 \label{3.6}
 \end{equation}
 with $g$ another constant. Inserting  the previous solution into (2.6) and (2.7) , one obtains
 \begin{equation}
 \sigma = - p = - \frac{g}{2 \pi}  
 \label{3.7}
 \end{equation}
 Keeping in mind that $\sigma < 0$ , we see that $g$, with dimension of $1/(distance)$, represents the acceleration of the domain wall, i.e. $g = a$. Therefore, our solution (3.6) is similar with that obtained by Visser \cite{MV} for the equation of motion of the classical membrane in the regime $R >>2 m$, when the spacetime is quasi Minkowskian. 
 
 To have a better look on the physical situation, we express the radius $R$ of the throat in terms of the Schwarzschild or Minkowski (for $R >>2 m$) time $t$. From (2.1) we have $dt/d \tau = cosh g\tau$, whence $t(\tau) = (1/g) sinhg\tau$ . One concludes that 
 \begin{equation}
 R^{2} - t^{2} = \frac{1}{g^{2}} ,
 \label{3.8}
 \end{equation}
 which is the equation of motion of the throat viewed by a Minkowskian observer at $R >>2 m$. It is nothing else but (2.8), with the acceleration $g = - 2 \pi \sigma > 0$. 
 
 The solution (3.6) is, of course, unstable and the wormhole blows up hyperbolically at infinity. We see from (3.7) that , although the throat is expanding, $\sigma$ and the surface tension $-p$ are constant ( see \cite{TD} where the author supposed a membrane model for the black hole horizon viewed as a ''viscous fluid bubble'' with a surface pressure $ g/8 \pi$ , where his $g$ is the black hole surface gravity ; see also \cite{PW} ). 
 
 For the scalar expansion of the throat we have
 \begin{equation}
 \Theta = \nabla_{\alpha} u^{\alpha} = 3 g ,
 \label{3.9}
 \end{equation}
 with $\dot{\Theta} \equiv u^{\alpha} \nabla_{\alpha} \Theta = 0$. In addition, the trace of the extrinsic curvature tensor is $K \equiv K^{i}_{i} = 3g$ , where the expression (3.6) for $R(\tau)$ has been used. The fact that $\Theta$ is nonvanishing means that the fluid should have some bulk viscosity $\zeta$ \cite{PW} \cite {AN} \cite{HC2}. If we suppose that our pressure $p$ is given only by the viscous term $- \zeta \Theta$, one obtains
 \begin{equation}
 \frac{g}{2 \pi} = - \zeta \Theta .
 \label{3.10}
 \end{equation}
 Using now (3.9) we get a negative bulk viscosity coefficient $\zeta = - 1/6 \pi$. That is not surprising as Parikh and Wilczek \cite{PW} obtained a similar result, explaining its negative value through the instability of the system.

\section{Conclusions} 
 By surgically grafting two Schwarzschild spacetimes together at $r>2m$ we get a manifold with no event horizon which has two asymptotically flat regions connected by a wormhole. We found its throat is expanding hyperbolically, with an acceleration proportional to the surface energy dnsity $\sigma$ on the throat. 
 
 We conjectured a classical membrane model for the wormhole with the equation of state $\sigma + p = 0$. The jump in the extrinsic curvature across the hypersurface $r = R(t)$ leads to a stress tensor corresponding to an expanding fluid with a negative bulk viscosity $\zeta = - 1/6 \pi$.

\end{document}